\documentclass[american,english,notitlepage]{revtex4-1}
\usepackage[T1]{fontenc}
\usepackage[utf8]{inputenc}
\usepackage{float}
\usepackage{units}
\usepackage{textcomp}
\usepackage{amssymb}
\usepackage{graphicx}
\usepackage{babel}
\begin{document}

\title{High-order parametric generation of coherent XUV radiation}
\selectlanguage{american}%

\author{O.\foreignlanguage{english}{ Hort }\textsuperscript{\selectlanguage{english}%
1,2,3\selectlanguage{american}%
}\foreignlanguage{english}{, }}
\email{Ondrej.Hort@eli-beams.eu }

\author{A. Dubrouil \textsuperscript{1,\#}, M.A. Khokhlova \textsuperscript{4,5},
D. Descamps \textsuperscript{1}, \foreignlanguage{american}{S.} Petit
\textsuperscript{1}, \foreignlanguage{american}{F.} Burgy \textsuperscript{1},
E. Mével \textsuperscript{1}, E. Constant \textsuperscript{1,6}
and V. V. Strelkov \textsuperscript{5,7}}

\affiliation{\textsuperscript{1} Université de Bordeaux, CNRS, CEA, Centre Laser
Intenses et Applications (CELIA), 43 rue P. Noailles, 33400 Talence,
France }

\affiliation{\textsuperscript{2} Photonics Institute, Vienna University of Technology,
Gusshausstrasse 27, A-1040 Vienna, Austria }

\affiliation{\textsuperscript{3} ELI Beamlines, FZU -- Institute of Physics
of the Czech Academy of Sciences, Na Slovance 1999/2, 182 21 Praha
8, Czech Republic }

\affiliation{\textsuperscript{4} Blackett Laboratory, Imperial College London,
London SW7 2AZ, United Kingdom }

\affiliation{\textsuperscript{5} Prokhorov General Physics Institute of the Russian
Academy of Sciences, 38 Vavilova st., Moscow, 119991, Russia }

\affiliation{\textsuperscript{6} Université de Lyon, Université Claude Bernard
Lyon 1, CNRS, Institut Lumière Matière (ILM), 69622 Villeurbanne,
France }

\affiliation{\textsuperscript{7} Moscow Institute of Physics and Technology (State
University), Dolgoprudny, Moscow Region, 141700, Russia}

\affiliation{\# Current Address: Femtoeasy, Femto Easy SAS, Bât. Gienah, Cité
de la Photonique, 11 avenue de Canteranne, 33600 Pessac, France}

\maketitle
\textbf{We observe a new regime of coherent XUV radiation generation
in noble gases induced by femtosecond pulses at very high intensities.
This XUV emission has both a reduced divergence and spectral width
as compared to high-order harmonic generation (HHG). It is not emitted
at a moderate intensity of the driving pulses where only high-order
harmonics are generated. At high driving intensities, the additional
XUV comb appears near all harmonic orders and even exceeds the HHG
signal on the axis. The peaks are observed in several gases and their
frequencies do not depend on the driving intensity or gas pressure.
We analyze the divergence, spectral width and spectral shift of this
XUV emission. We show that these specific features are well explained
by high-order parametric generation (HPG) involving multiphoton absorption
and combined emission of an idler THz radiation and an XUV beam with
remarkably smooth spatial and spectral characteristics.}

\section*{Introduction}

In recent decades, X-ray free-electron lasers, synchrotrons, and laser-driven
XUV and X-ray sources, have become the most widespread generators
of coherent and ultrashort pulse radiation in this spectral range
\citep{Rocca1999,Daido2002,Krausz_2009,McNeil_2010}, and are used
for various applications in physical chemistry, atomic physics and
coherent imaging \citep{Leone2016,Yun_2017,Chapman_2010}. Among the
laser-driven sources, high-harmonic generation (HHG) in gases achieves
very good spatial and temporal coherence, stability and setup compactness
when compared to X-ray free-electron lasers and synchrotrons \citep{McPherson1987,Ferray1988,Corkum1993,Balcou1997,Constant1999}.
However, HHG suffers from poor conversion efficiency resulting in
low XUV photons number. This limits the interdisciplinary spread of
XUV applications.

The effort to increase the generated XUV signal resulted in number
of methods addressing amelioration of the phase matching, exploiting
quasi-phase-matching, low-order wave mixing or even XUV amplification
\citep{Rundquist1998,Zhang2007,Misoguti_2005,Bredtmann_2017}. Different
way is being opened with high energy Ti:Sapphire and OPCPA-based laser
systems where high XUV photon number is obtained even with state-of-the-art
HHG conversion efficiency \citep{Nayak2018,Hort2019}.

Here we present the experimental and theoretical study of the high
driving intensity regime of HHG and observed parametric generation
of XUV radiation. We found that this new XUV source signal keeps rising
at driving intensities where standard HHG saturates. Therefore, it
may represent a way to significantly increase XUV signal for photon-hungry
applications across various science fields.

\section*{Results}

\subsection*{Experimental observation of parametric XUV signal}

\begin{figure}
\begin{centering}
\includegraphics{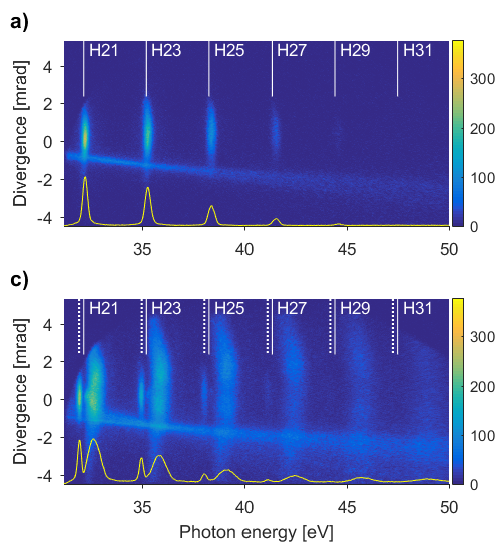}\includegraphics{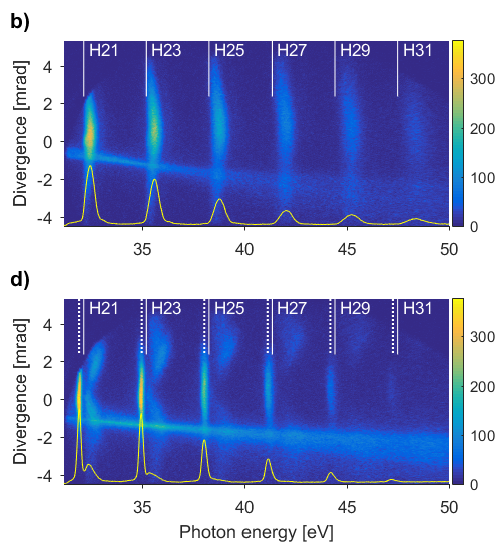}
\par\end{centering}
\caption{\textbf{Experimental spatially resolved XUV spectra generated in a
krypton jet.} The colorbar shows spatio-spectral intensity (in arbitrary
units). The reference photon energy $q\omega_{0}$ of the high-order
harmonics (generated at low intensity) is presented by the solid white
lines, while the photon energy of the red-side peaks given by equation
(\ref{eq:omega2}) with the parameter $Q=27$ and $m=2$ (see below)
are shown by dashed lines. Solid yellow lines present integrated on-axis
XUV signal in the full angle of 1.5 mrad (plotted in arbitrary units
and arbitrary offset). The driving laser intensity (estimated for
propagation under vacuum) is (from a to d) 0.15, 0.71, 2.9, and $\unit[3.5\times10^{15}]{Wcm^{-2}}$.
\label{fig: HHG Spectra Krypton-Energy}}
\end{figure}

We generate XUV radiation in gases with a Ti:Sapphire laser in a loose
focusing geometry at high intensity and characterize, spatially and
spectrally, the emitted XUV light. We use multi-mJ TW pulses of $\unit[45]{fs}$
duration centered at 810 nm with a spatially-filtered beam (see Methods).
Figure \ref{fig: HHG Spectra Krypton-Energy} shows the XUV spectra
obtained in krypton gas jet at different driving laser intensities.
Conventional high-order harmonics (HH) appear at low intensity and
red-side satellites (RSS) appear on the low-frequency side of the
harmonic peaks when high driving intensities are used (see Supplementary
Fig. S2 for the full dataset).

At a low driving intensity of $\unit[0.15\times10^{15}]{Wcm^{-2}}$
(Fig. \ref{fig: HHG Spectra Krypton-Energy} a), the harmonics are
spectrally symmetric with a low spatial divergence and the peaks are
located at photon energy $q\omega_{0}$ with only a small blue shift.
The highest generated harmonic is H29. With a higher driving intensity
(Fig. \ref{fig: HHG Spectra Krypton-Energy} b) the cutoff rises and
there is a pronounced blue shift and spectral asymmetry of the harmonics.
The asymmetry is due to the ionization of the generating medium that
confines HHG in the pulse rising front and the blue shift is due to
IR pulse propagation in the ionized medium that shifts the laser central
frequency. Simultaneously the XUV beam divergence increases with driving
intensity.

At high driving intensity of $\unit[2.9\times10^{15}]{Wcm^{-2}}$(Fig.
\ref{fig: HHG Spectra Krypton-Energy} c) additional peaks, here referred
to as RSS, appear on the red side of harmonics. At very high driving
intensity (Fig. \ref{fig: HHG Spectra Krypton-Energy} d) the RSS
signal further increases. All RSS peaks are red-shifted as compared
to the spectral position of the harmonics generated at very low laser
intensities and exhibit a lower spatial divergence. In general, with
increasing the driving intensity, additional RSS appears near higher
order harmonics. This behavior presents itself as a cut-off that rises
with the driving laser intensity. 

\begin{figure}
\begin{centering}
\includegraphics{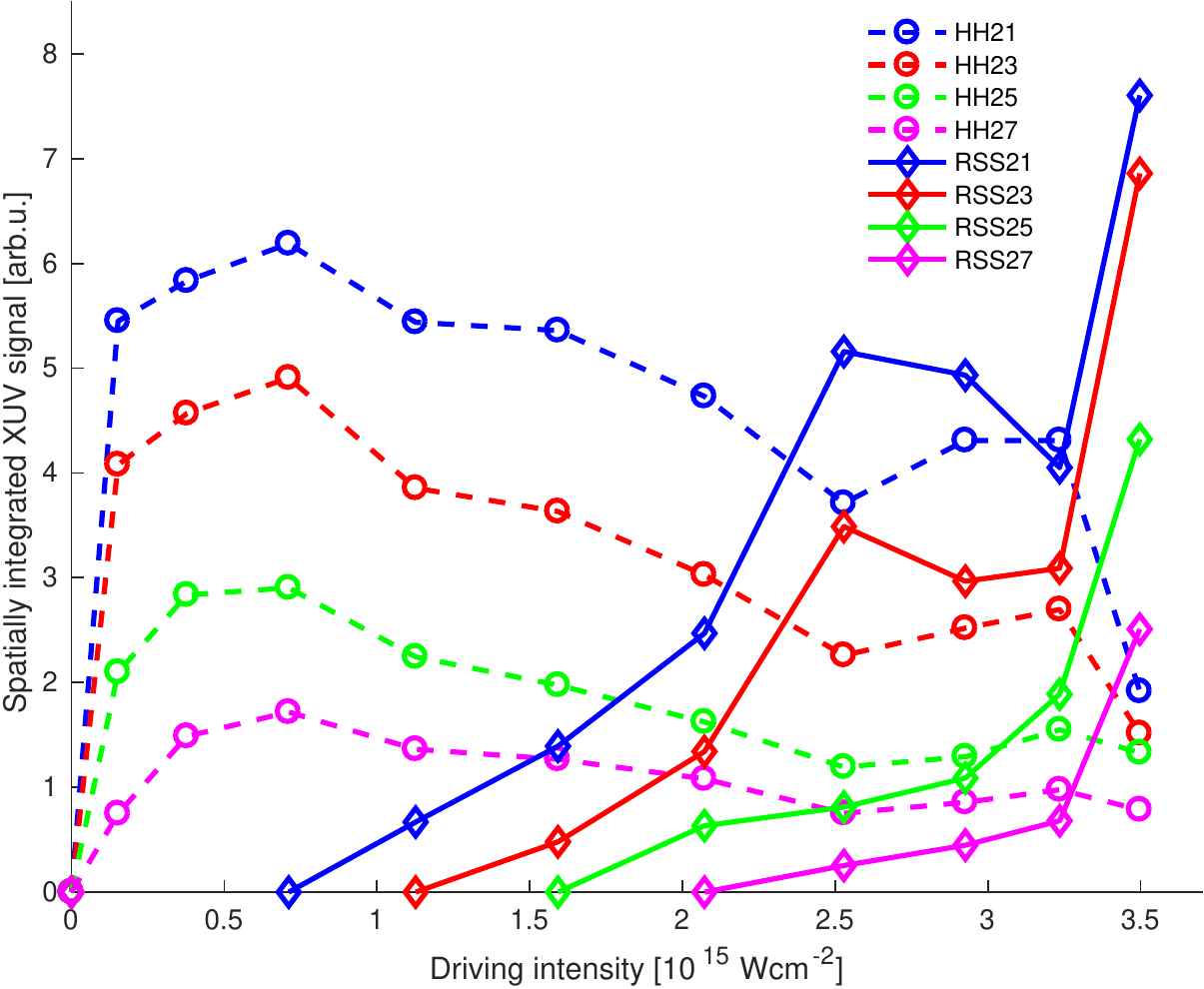}
\par\end{centering}
\caption{\textbf{Measured on-axis XUV signal of orders 21 - 27.} Experimental
comparison of the spatially integrated on-axis XUV signal (in the
full angle of 0.5 mrad) of the HH (dashed circles) and RSS peaks (solid
diamonds) as a function of the driving laser intensity.\label{fig:Signal}}
\end{figure}

The HHG and RSS signals evolve in different ways. To illustrate this,
Fig. \ref{fig:Signal} shows the measured XUV signal on axis as a
function of driving intensity corresponding to Fig. \ref{fig: HHG Spectra Krypton-Energy}. 

With increasing driving intensity the HH signal rises up to saturation
at $\unit[0.7\times10^{15}]{Wcm^{-2}}$ and then starts to decrease
slowly. This is attributed to a degradation of the phase-matching
conditions for HHG and high ionization of the medium modulating the
spatio-temporal profile of the IR pulse \citep{Gaarde2008,Dubrouil2014}.
In contrast, the RSS signal first appears only around $\unit[1.2\times10^{15}]{Wcm^{-2}}$
and \emph{continues to grow} with an increase in driving intensity.
It does not reach saturation and its evolution is qualitatively very
different to HH. At intensities above $\unit[3.25\times10^{15}]{Wcm^{-2}}$
the RSS peaks are of similar or higher on-axis brightness than those
of the HH. As mentioned above, there is a threshold driving intensity
for every RSS order that increases with the order (RSS cutoff). 
\begin{center}
\begin{figure}[H]
\begin{centering}
\includegraphics{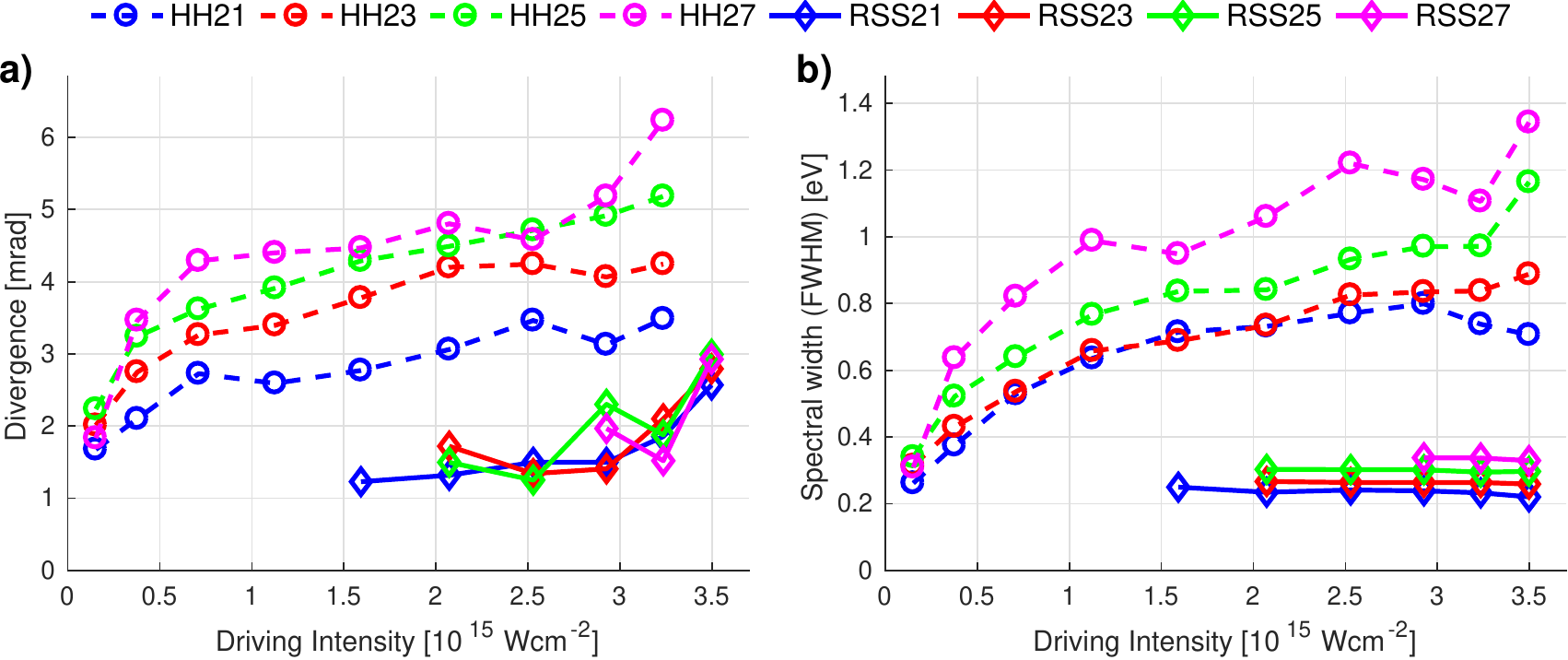}
\par\end{centering}
\caption{\textbf{Measured divergence and spectral width of the XUV beam.} Divergence
(a) and spectral width (b) measured at FWHM of the HH (dashed circles)
and RSS (solid diamonds) as a function of the driving laser intensity.
\label{fig:Divergence_spectral_width}}
\end{figure}
\par\end{center}

The spatial profile and spectral width of the RSS are very different
to the ones of the HH beam. Figure \ref{fig:Divergence_spectral_width}
a presents the spatial divergence of the HH and RSS beams as a function
of the driving laser intensity. The HH divergence increases quickly
with intensity up to $\unit[0.7\times10^{15}]{Wcm^{-2}}$. Then its
spatial profile keeps widening up to the high intensity of $\unit[3.5\times10^{15}]{Wcm^{-2}}$
where it becomes very irregular, with its size comparable to the 40
mm MCP detector diameter. On the contrary, the divergence of the RSS
beam is lower than the HH beam and does not change much with the driving
laser intensity over a large range of intensities up to $\unit[2.9\times10^{15}]{Wcm^{-2}}$.
Above this, the RSS beam expands while keeping its spatial profile
regular.

Figure \ref{fig:Divergence_spectral_width} b shows the spectral width
corresponding to Fig. \ref{fig: HHG Spectra Krypton-Energy}. The
spectral width of the RSS is close to that of the HH for the lowest
driving intensity, and significantly lower for the higher ones. Moreover,
the RSS frequency does not shift or broaden significantly with the
increase of the driving intensity, even when the gas medium becomes
strongly ionized.

From the Figs. \ref{fig: HHG Spectra Krypton-Energy}, \ref{fig:Signal}
and \ref{fig:Divergence_spectral_width} it is clear that the HH and
RSS peaks behave differently. Moreover, the spectrum and the beam
shape of the RSS radiation do not depend on the gas pressure in the
generating medium while HH do (Fig. S1 in Supplementary). This implies
that HH and RSS originate from different processes. 

\subsection*{Comparing of the observed features to the state-of-the-art}

Many nonlinear processes occurring in a gas jet irradiated by a strong
laser field leading to XUV light emission have already been reported.
They all exhibit specific features that are not observed here and
they cannot explain our observations. In the following sections, we
compare our experimental results with published findings and briefly
present the theory which explains the high-order parametric generation
(HPG) process that has been recently proposed and show that this process
fits well our observations.

The XUV spatio-spectral shape has been studied already in several
papers \citep{Kan1995a,Zhou1996,Zair2008,Brunetti2008,Cao2012,He2015,Catoire2016,Zhang_2019,Strelkov2020}
showing complex harmonic line broadening and splitting with an increase
of the driving laser intensity. These studies demonstrated (i) \emph{continuously}
evolving XUV spectrum with increasing laser intensity and (ii) more
pronounced spectral features for the lower harmonics and \emph{progressive}
disappearance of features for the higher ones. This behavior is very
different from our observations, demonstrating the sudden appearance
of the red-side peaks for high laser intensity, no pronounced evolution
of their spectral frequency under further intensity increase and similar
spectral frequency in the plateau and cut-off regions. Therefore,
the explanation offered in \citep{Kan1995a,Zhou1996,Zair2008,Brunetti2008,Cao2012,He2015,Catoire2016,Zhang_2019}
does not fit our case. Moreover, the intensities that we use here
with Kr are higher than those used in the former studies and allow
accessing new phenomena.

Another group of experiments reported the modification of the XUV
spectrum due to resonances of the generating particles \citep{Wang2010,Shiner2011,Ganeev2012,Beaulieu2016,Bengtsson2017,Fareed2017,Beaulieu2017}.
However, the resonance-induced modification of the spectrum is specific
for each type of generating particles. In contrast to this, in our
experiments spectral frequency of the RSS is robust and similar for
different gases (Figs. S2-6 in Supplementary). Also, the value of
the RSS spectral frequency changes with the gas target geometry (jet
or capillary) of the same gas. So the observed spectral features cannot
be attributed to resonances.

The publications reporting the amplification of XUV were focused on
stimulated emission via population inversion \citep{Serrat2013,Bredtmann_2017}
or other approaches using two-color \citep{Reinhardt2000,Dao2015}
or single-color \citep{Seres2010} driving field. In these papers
the emission on HHG frequencies (or at corresponding combination of
frequencies for the two-color case) were studied and the emission
of \emph{new} spectral components were not reported, in contrast to
this paper.

In our case, there is no signal at even harmonic orders. This leads
to exclusion of various low-order wave mixing such as sum- or difference-frequency
generation \citep{Misoguti_2005}, prepulses or chirp of the driving
pulse. Figure \ref{fig:Schematic-comparison} b also illustrates that
the position of the RSS does not correspond to odd multiple of any
infrared frequency and cannot therefore be due to HHG with a spectrally
shifted pre- or post-pulse.

\subsection*{General properties of High-Order Parametric Generation}

\begin{figure}
\begin{centering}
\includegraphics{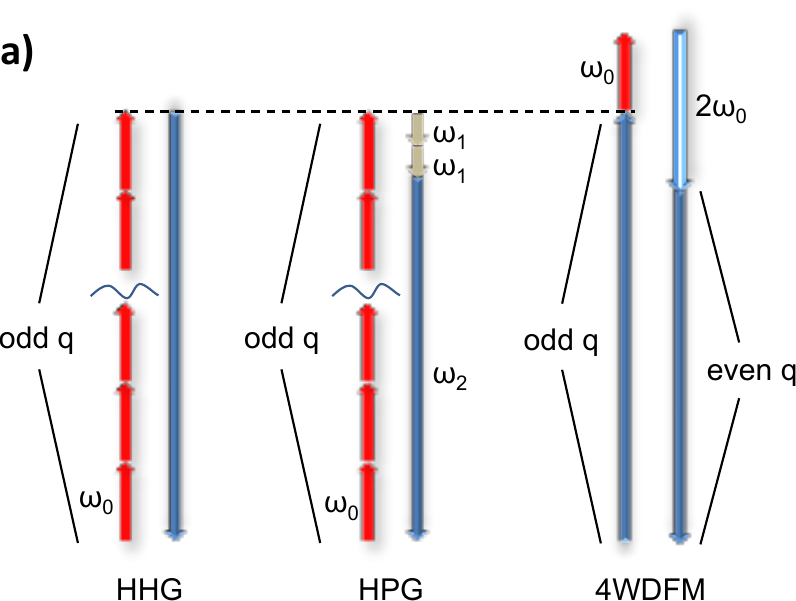}\includegraphics{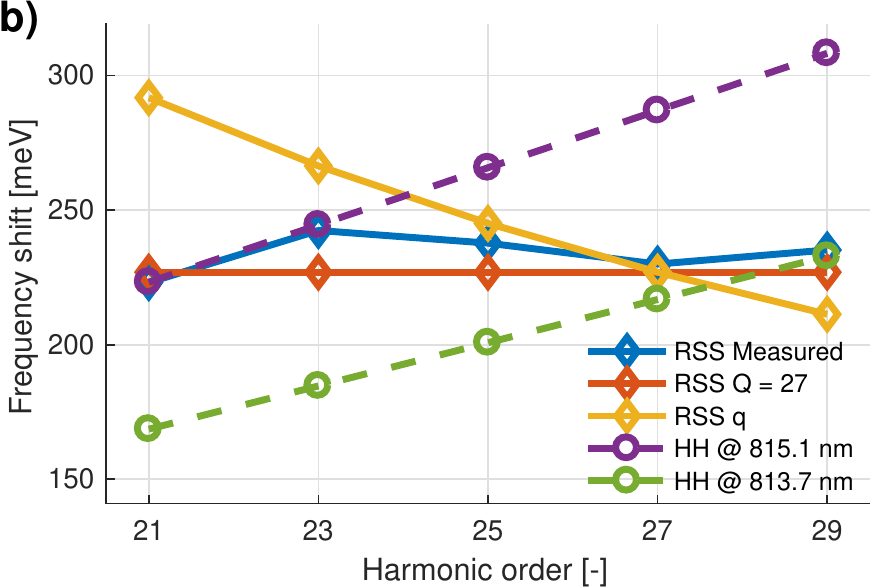}
\par\end{centering}
\caption{\textbf{Schematic comparison of (a) nonlinear processes that lead
to coherent XUV emission and (b) measured RSS photon energies with
calculated values: }(a) High-order harmonic generation, High-order
parametric generation (with $m=-2$) and four-wave difference-frequency
mixing \citep{Misoguti_2005} (b) The measured data (solid blue diamonds)
are compared to values calculated with Q = 27 (solid red diamonds),
q equal to the harmonic order (solid yellow diamonds), high harmonics
corresponding to the driving central frequency of 815.1 nm (dashed
purple circles) and 813.7 nm (dashed green circles).\label{fig:Schematic-comparison}}

\end{figure}

The features observed in our experiments are consistent with the HPG
that was recently predicted by V. Strelkov \citep{Strelkov2016}.
This process is analogous to well-known (low-order) parametric generation
but it involves many laser photons corresponding to the intense-field
domain. While in the HHG several laser photons turn into a single
XUV photon, in the high-order parametric process they turn into few
photons generated in a different spectral range (XUV to THz), see
Fig. \ref{fig:Schematic-comparison}.

The nature of the nonlinearity leading to HHG and HPG is the same,
just as low-order nonlinearity of a crystal can lead to low-order
harmonic generation or parametric generation, depending on the experimental
conditions. In our case the origin of the high-order nonlinearity
is a rescattering process \citep{Corkum1993} so both HHG and HPG
take place at similar laser intensities above a threshold driving
intensity. This threshold is a consequence of minimal number of IR
photons needed for the process. 

Similar to low-order parametric generation, the value of the generated
frequencies of idler and signal is defined by phase-matching conditions.
In our case the parametric process leads to generation of the XUV
photons with lower photon energy than the HH photons and therefore
these appear at the red side of the harmonics in the spectrum. The
frequency $\omega_{2}$ of this parametric signal is: 
\begin{equation}
\omega_{2}=q\omega_{0}+m\omega_{1}\,,\label{eq:omega2}
\end{equation}

where $\omega_{0}$ is the driving laser frequency, $q$ is the harmonic
order, $m$ is a negative low even number and $\omega_{1}$ denotes
the idler frequency. 

In \citep{Strelkov2016} it is shown that the process is efficient,
assuming that at high driving intensity the plasma dispersion provides
the dominant contribution to the phase mismatch. In that case, the
detuning from the phase-matching condition is $\ensuremath{\Delta k=-\frac{\omega_{pl}^{2}}{2c}(\frac{q}{\omega_{0}}+\frac{m}{\omega_{1}}-\frac{1}{q\omega_{0}+m\omega_{1}})}$,
where $\ensuremath{\omega_{pl}}$ is the plasma frequency, defined
by the density of the free electrons in the generating gas. Thus for
the idler frequency $\omega_{1}=-\frac{\omega_{0}m}{q}$ the HPG process
is almost phase-matched \emph{regardless of the electronic density}
(see Fig.\ref{fig:Phasematching}). This is very important because
the latter naturally varies in space and time during the generation,
and any phase mismatch significantly limits coherent XUV emission
via HHG. We illustrate this in Fig. \ref{fig:Signal}, where the HH
signal decreases with increasing IR intensity, and the RSS signal
increases as a consequence of the phase-matched parametric process. 
\begin{center}
\begin{figure}[H]
\begin{centering}
\includegraphics{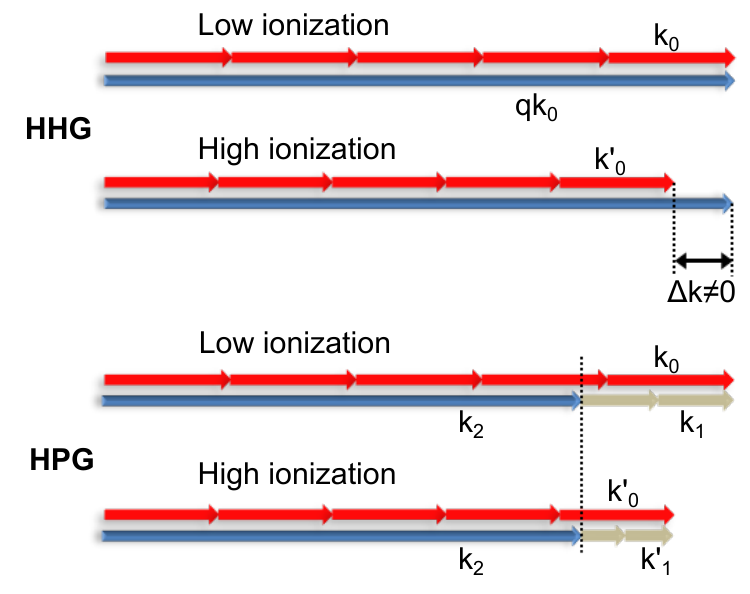}
\par\end{centering}
\caption{\textbf{Schematic comparison of the HHG and HPG phase-matching and
blue shift with the presence of low and high ionization.} The high
electronic density creates a phase mismatch $\Delta k\protect\neq0$
in case of HHG. In the case of HPG, the change of the fundamental
$k_{0}$ is compensated by the change of $k_{1}$ so $\Delta k=0$
is valid even under high ionization and the RSS (wave number $k_{2}$)
are phase-matched and keep their spectral frequency.\label{fig:Phasematching}}
\end{figure}
\par\end{center}

For the lowest order ($m=-2$) of the idler emission, $\omega_{1}=\frac{2\omega_{0}}{Q}$
(with $Q\thickapprox q$) and it can contribute to the phase-matched
emission of RSS frequency $\omega_{2}$ given by equation (\ref{eq:omega2}).
Fig. \ref{fig: HHG Spectra Krypton-Energy} shows that this equation
describes the HPG spectral frequency very well, assuming the lowest
possible order $m=-2$ and choosing the order $Q$ close to the mean
value of the harmonic order in the plateau region. We stress here
that the parameters $m$ and $Q$ cannot be chosen arbitrarily. Indeed,
$m=-2$ is the most favorable for krypton and argon gas medium, as
a parametric process in general is less efficient for higher parametric
orders (i.e. higher absolute values of $m$). Our approach is similar
towards the parameter $Q$. In \citep{Strelkov2016} it is considered
that every single harmonic has its own RSS satellite, generated via
specific $\omega_{1}$. However, frequencies $\omega_{1}=\frac{2\omega_{0}}{q}$
are very close to each other for neighbor harmonics. We assume that
the identical idler frequency $\omega_{1}=\frac{2\omega_{0}}{Q}$
is generated for all plateau orders. In this case all the harmonic
orders would contribute to the generation of this frequency $\omega_{1}$,
so the parametric process becomes more efficient.

In our conditions the idler frequency $\omega_{1}$ is in the spectral
range of few tens of THz (27 THz corresponding for Fig. \ref{fig: HHG Spectra Krypton-Energy}).
Such kind of radiation can be created directly in the gas jet during
the gas ionization. It was already successfully generated in plasma
produced by focusing high intensity femtosecond pulses into a gas
target. Mixing of the fundamental and second harmonic of the laser
is widely used to efficiently generate THz radiation \citep{Kim2008,Clerici2013,Vvedenskii2014,Andreeva2016}
and fundamental frequency only produces THz radiation with about one
or two orders of magnitude lower efficiency \citep{Hamster,Roskos_2007,DAmico2007,Bree2017}.

\subsection*{Comparison between experimental observations and theoretical predictions
for HPG}

\subsubsection*{Frequencies of the RSS spectral peaks}

The HPG theoretical predictions are highly consistent with the experimental
results. Remarkably, in contrast to published works (mentioned above),
it produces spectral peaks for all the harmonic orders. In Fig. \ref{fig:Schematic-comparison}
b we compare our experimentally measured RSS spectral frequency to
the calculated values using different methods and different fundamental
wavelengths. As explained above, using one $Q$ parameter for neighboring
harmonics leads to excellent agreement between the calculated and
measured data. In contrast to this, the possibility of RSS being generated
by other processes such as HHG of longer fundamental wavelength is
ruled out.

\subsubsection*{Widths of the RSS spectral peaks}

HPG also explains why the RSS spectral width is much lower than that
of the HH. The RSS are spectrally narrow because the parametric signal
enhancement is, in particular, proportional to its intensity, so the
most intense frequency component is the most enhanced. Moreover, the
blue shift acquired by the fundamental wave $\omega_{0}$ during ionization
is compensated by the blue shift of $\omega_{1}$ during ionization,
so the RSS spectrum centered at $\omega_{2}$ does not shift with
increase of driving laser intensity. To explain this in more detail
let us denote the ionization-induced blue-shift of the fundamental
as $\Delta\omega_{0}$; the blue shift of the low-frequency (idler)
field $\omega_{1}$ is inversely proportional to its frequency: $\Delta\omega_{1}=\frac{\Delta\omega_{0}Q}{|m|}$.
So for the polarization response at the RSS frequency given by equation
(\ref{eq:omega2}) the blue shifts of the two generating waves compensate
each other: $\Delta\omega_{2}=q\Delta\omega_{0}+m\frac{\Delta\omega_{0}Q}{|m|}\approx0$
(this compensation is similar to the phase mismatch compensation illustrated
in Fig. \ref{fig:Phasematching}). Opposite to this, the ionization-induced
blue shift of the fundamental leads to the pronounced blue shift of
the polarization response for the q-th HH, equal to $q\Delta\omega_{0}$.
Thus, the spectral broadening due to the blue shift is much more pronounced
for the HH than for the RSS. 

There is another mechanism of the spectral shift of the XUV generated
due to HHG and HPG, namely, the polarization response phase dependence
on the generating field(s) intensity. For the HHG this leads to the
blue shift at the rising front of the pulse and the red shift at the
falling front; however, the ionization temporally confines HHG to
the rising front, so this blue-shift adds to the ionization-induced
one, leading to a strong broadening of the HH. The HPG process is
phase-matched even for a high ionization degree of the gas, so it
takes place mainly near the maximum of the fundamental pulse where
the temporal variation of the intensity vanishes. Thus the polarization
phase dependence at this intensity does not lead to a pronounced frequency
shift or broadening of the RSS. 

The spectral width of the RSS is thus close to the inverse of the
pulse duration, which does not change much with the fundamental intensity.
In contrast to this, the HH is broadened due to the two mechanisms
described above, and this broadening increases in line with the fundamental
intensity. This considerations explain the experimental results presented
in Fig. \ref{fig:Divergence_spectral_width} b. The observed RSS spectral
width corresponds to a transform-limited XUV pulse duration of approximately
7.7, 6.9, 6 and 5.4 fs for RSS orders 21, 23, 25 and 27 respectively,
which is shorter than 45 fs of the fundamental pulse, as expected
for a high-order nonlinear process.

\subsubsection*{Angular divergence of the XUV}

The parametric XUV generation process also explains the narrow angular
divergence of the XUV emission observed experimentally as shown in
Fig. \ref{fig:Divergence_spectral_width}. 

As discussed above, the generation of RSS is phase-matched regardless
of the electronic density (in contrast to HHG, see Fig. \ref{fig:Phasematching}).
The RSS are therefore efficiently generated both on the optical axis
and periphery of the fundamental beam resulting in large XUV beam
size and low beam divergence in the far field, irrespective of the
IR intensity. Note that, at low IR intensity when HHG also takes place
on axis, the HH divergence is similar to that of RSS (see Fig. \ref{fig:Divergence_spectral_width}).

\subsection*{Numerical study}

As was shown in a number of papers (for a review see \citep{Roskos_2007}),
an intense few-cycle laser pulse can generate a THz field because
the ionization takes place at few half-cycles and electronic motion
after ionization can be asymmetric. However, similar mechanism can
be valid for a multi-cycle pulse as well if it is very intense (namely,
its peak intensity is much higher than the photoionization threshold
intensity): the ionization takes place rapidly yet at the front of
the pulse, thus the high ionization degree can occur during few half-cycles.
To study this process we solve numerically 3D time-dependent Schrödinger
equation (TDSE) for an atom in external laser fundamental field and
calculate the spectrum of the microscopic response (see Methods section
for more details). When the fundamental intensity is so high that
the ionization occurs during several half-cycles, a continuum spectrum
in the multi-THz domain appears. While the microscopic response increases
with the frequency decrease in the range of few tens of THz, frequencies
below the plasma frequency ($\unit[15]{THz}$ for plasma density $\unit[3\times10^{18}]{cm^{-3}}$)
cannot propagate in the plasma. So the spectrum of the THz field propagating
in the target has a maximum $\unit[20-30]{THz}$. Our calculations
for argon show that the intensity of this THz field grows very rapidly
with the laser intensity when the latter is about $\unit[0.4\times10^{15}]{Wcm^{-2}}$
and then saturates. 

Solving the propagation equation we find the macroscopic response
of the medium. For our conditions the intensity of the $2/27\omega_{0}=\unit[27]{THz}$
radiation after propagation of half target length achieves level of
$\unit[10^{11}]{Wcm^{-2}}$. Note that such intensities of the THz
``seeding'' are orders of magnitude higher than the ones used in
\citep{Strelkov2016}. So the HPG can take place for much shorter
propagation distances and lower pressures, as confirmed by our results.

\begin{figure}[H]
\begin{centering}
\includegraphics[scale=0.6]{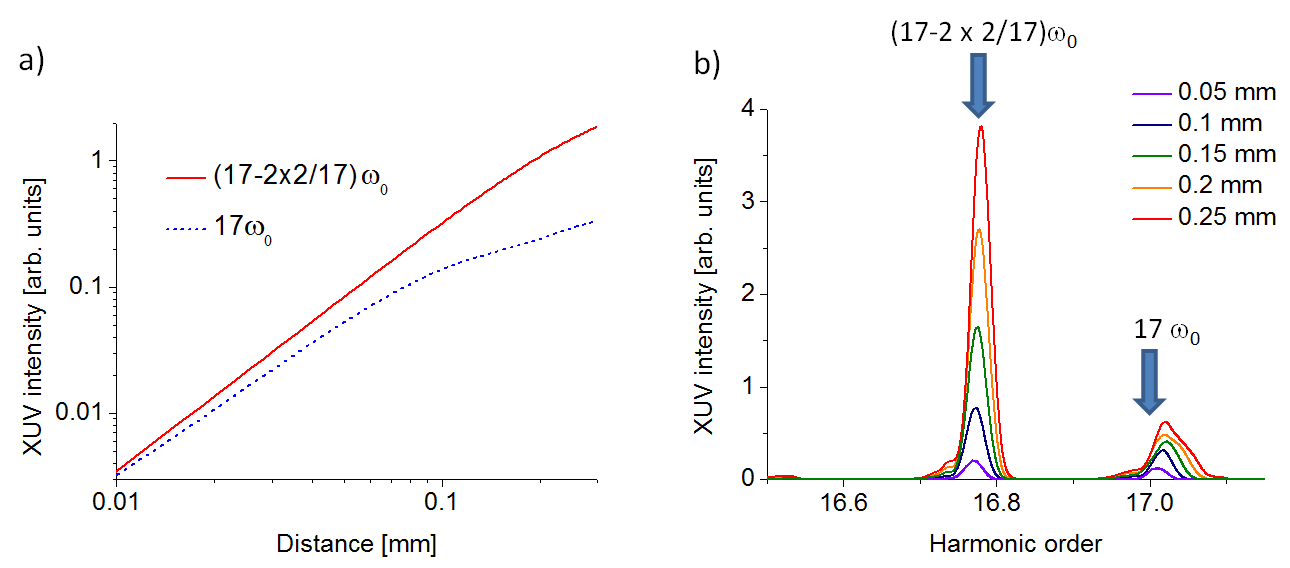}
\par\end{centering}
\caption{(a) Calculated intensities of the 17\protect\textsuperscript{th}
harmonic and the RSS (shown in the graph) as a function of the propagation
distance in log-log scale. (b) Spectrum near 17\protect\textsuperscript{th}
harmonic for several propagation distances. Calculation is done for
argon, the gas density is $\unit[3\times10^{18}]{cm^{-3}}$, laser
intensity is $\unit[0.2\times10^{15}]{Wcm^{-2}}$\label{fig:calc_results}}

\end{figure}

In our further calculations we study the XUV generation by the laser
field and the THz field. Figure \ref{fig:calc_results} a shows the
intensity of the harmonic and RSS as a function of the propagation
distance. We see that the harmonic intensity behaves in agreement
with numerous HHG studies: first it grows quadratically and then saturates
(note the log-log scale). However, the intensity of the RSS keeps
almost quadratic growth through the full length of the simulation
because of much better phase-matching for the generation of the RSS.
So after the propagation the RSS becomes significantly more intense
than the harmonic. That corresponds to our experimental data on Fig.
\ref{fig:Signal}. 

Figure \ref{fig:calc_results} b presents the generated spectrum near
the 17\textsuperscript{th} harmonic. One can see that the RSS line
is narrower than the the harmonic line in agreement with Fig. \ref{fig:Divergence_spectral_width}
b. The origin of this narrowing for the process of HPG was discussed
above. 

\section*{Discussion}

To the best of our knowledge, the HPG-like XUV spectra were not reported,
although our experimental conditions are within the reach of current
technology. We believe that it is caused by the unique combination
of our experimental parameters together with the fact, that we extensively
study the experimental conditions that are believed to be inefficient
for HHG. Our generating conditions are very specific in a way that
the short medium in the high ionization regime is far above phase-matched
HHG. Note, that the RSS can easily be misinterpreted with alignment
problems or IR spectral splitting. However, our results show that
the RSS photon energy does not fit a multiple of any fundamental photon
energy, therefore HHG emitted by a chirped pulse or by a prepulse
are ruled out.

In conclusion, we have observed spectrally narrow, low divergence
XUV emission at high intensity driving pulses and we show that all
features of this emission are consistent with high-order parametric
emission where an intense laser pulse can generate XUV and THz photons
in a phase-matched process irrespective of the electron density. The
observed parametric signal can outreach the on-axis signal of regular
HHG because it rises at driving intensities where the HHG signal saturates,
giving the possibility of upscaling. In this sense, with high energy/high
peak power laser system, the HPG can produce XUV radiation with higher
brightness and intensity than HHG. Moreover, as the XUV spectral bandwidth
of the RSS is of similar order as the HH, we expect that one can generate
attosecond pulses via HPG. 

Furthermore, this XUV parametric process may be improved. In analogy
with the better understood parametric processes in the visible and
mid-IR range, we anticipate that controlling the idler will provide
numerous ways to enhance and control temporally and spatially this
bright coherent XUV beam. Further increasing of the driving pulse
energy/intensity can be used to prove the scalability and complete
temporal characterization could give more insight into the process
of HPG. Finally, as the spectro-spatial profile of the RSS is very
regular and does not depend on the driving pulse intensity and ionization
degree, the HPG can be more attractive than HHG even for applications
where focusability into very small spot is crucial.

\section*{Methods}

\subsection*{Experiment}

To perform the experiment we used a chirped-pulse amplification (CPA)
based Ti:Sapphire laser chain delivering multi-mJ TW pulses of $\unit[45]{fs}$
duration centered at $\unit[810]{nm}$ at a repetition rate of $\unit[10]{Hz}$.
The beam was spatially filtered by a hollow core capillary. The Strehl
ratio was measured by a wavefront sensor (HASO) giving the value of
0.95. The collimated IR beam with $w=\unit[10]{mm}$ radius at $\nicefrac{1}{e^{2}}$
was clipped by an iris of $\unit[22]{mm}$ diameter before being focused
by a mirror of focal length of $\unit[2]{m}$ into a gas jet with
a focus spot of $w=\unit[73]{\mu m}$ radius at $\nicefrac{1}{e^{2}}$
(corresponding to a Rayleigh length of $\unit[21]{mm}$). The nozzle
of the gas jet is $\unit[250]{\mu m}$ in diameter leading to a $\unit[0.8]{mm}$
medium length in the laser interaction region.The back pressure of
krypton gas was around $\unit[3]{bars}$. The medium length and the
pressure in the interaction zone were estimated to $\unit[100-150]{mbar}$
with a gas density measurement technique similar to \citep{Comby2018}.

The generated XUV radiation was then spectrally and spatially resolved
by a flat field spectrometer consisting of $\unit[0.5]{mm}$ entrance
slit imaged by a 3° grazing-incidence Hitachi concave grating ($\unit[1200]{gr/mm}$)
onto 40-mm-diameter dual multi-channel plates (MCP) coupled to a phosphor
screen. The XUV signal is therefore measured behind the slit that
is centered on the beam axis. All data presented are only corrected
from background signal of the detector. The horizontal continuous
line in the lower part of the krypton and argon spectra is caused
by a diffusive reflection of low harmonic orders in a vacuum tube
between the XUV grating and the MCP detector and therefore has no
significance for the results.

In such conditions the XUV radiation is generated in a loose-focusing
regime where the generating medium is much thinner than the Rayleigh
range of the driving beam. One should note that the XUV signal is
high enough to acquire the spatially resolved spectra on a single-shot
basis (see Fig. S3 and S6), although for better statistics the presented
spectra were acquired as 10 shots average.

We stress that using loose focusing and spatially filtered beam together
with acquiring high resolution spatially resolved spectra with wide
spectral range facilitate greatly the recognition of such phenomena
as HPG rather than HHG.

The values of IR pulse intensity are estimated for vacuum (linear)
propagation of the laser beam. During the laser pulse propagation
in the medium, when the gas is ionized, the rise in intensity is not
directly proportional due to the plasma defocusing and the peak intensity
can even decrease.

\subsection*{Simulation}

The microscopic response is calculated via numerical solution of the
3D time-dependent Schrödinger equation (TDSE) for a model single-active
electron atom in external field. This microscopic polarization is
used in the propagation equation to calculate the macroscopic response.

However, the full numerical integration of the 3D propagation equation
is very heavy. Consequently we split it into two parts. Within the
first part we study the THz field generation near the beam axis, using
very high peak intensities achieved on axis in our experiments. In
contrast to majority of the studies of the THz field generation, we
focus on the THz field \emph{inside} the generating medium. We solve
the 1D propagation equation: 
\[
\frac{\partial E_{\omega}(z)}{\partial z}=-\frac{i2\pi\omega}{c}P_{\omega}(z),
\]
 where $E_{\omega}$ is a field amplitude at the frequency $\ensuremath{\omega}$,
$\ensuremath{P_{\omega}}$ is the polarization of the medium which
is proportional to the gas density and the single-atom response calculated
via 3D TDSE (the numerical approach for the TDSE solution is described
in \citep{Strelkov2006}). The equation is solved numerically by a
slice-by-slice propagation: the field found at an n\textsuperscript{th}
slice is used in the TDSE solution to calculate the polarization and
thus the field at the next (n+1)\textsuperscript{th} slice (the details
of the numerical approach for the propagation equation integration
are described in \citep{Strelkov2020}). 

In these calculations we find that the THz radiation with frequencies
of about $\unit[20-30]{THz}$ can be efficiently generated in our
conditions.

At the second stage of our calculations we study XUV generation in
the laser and THz field. Complete simulation of our experimental conditions
exceeds our numerical capabilities, so we deal with shorter propagation
distances, lower harmonic orders and the THz field with somewhat higher
frequency of $2/17\omega_{0}$. Moreover, we consider the laser intensity
that is lower than at the beam axis, taking into account that the
periphery of the laser beam with such a level of intensity has relatively
high volume and that such intensities are favorable for HHG.

Note, that our approach of the macroscopic response calculation based
on polarization calculated via TDSE solution is exact (within 1D propagation
and single-active electron approximation), so it includes processes
of HHG, high-order wave mixing, HPG, plasma blue shift of the generating
and generated waves, and so on.

\bibliographystyle{unsrt}
\bibliography{HPG_Main}

\begin{acknowledgments}
We acknowledge financial supports from the CNRS, the RFBR (grant N
19-02-00739), the European commission (EU program Laserlab Europe
II and III), the ANR (ANR-09-BLAN-0031-02 Attowave), and in the frame
of the Investments for the future Programme IdEx Bordeaux---LAPHIA
(ANR-10-IDEX-03-02) and the region Aquitaine (COLA2 2-1-3-09010502
and NASA 20101304005). O.H. acknowledges financial supports from Ministry
of Education, Youth and Sport of the Czech Republic; European Regional
Development Fund (ERDF), European Social Fund (ESF) and European Union’s
Horizon 2020 research and innovation programme under grant agreements
No 657272 and 797688. M.K. acknowledges financial supports from The
Royal Society through Newton International Fellowship (NF161013).
E.C. acknowledges stimulating discussions with Stefan Skupin. V.S.
acknowledges stimulating discussions with Sergey Popruzhenko. The
numerical studies were funded by RSF (grant N 16-12-10279). We are
grateful to Rachael Jack for manuscript language revision.
\end{acknowledgments}

\section*{Author Contributions}

O.H., A.D., and E.C. built the experimental setup. O.H. performed
the experiment. O.H., V.S. and E.C. analysed the data. V.S. and M.K.
performed the calculations. S.P. built the laser system. S.P., D.D.
and F.B. managed the laser system. O.H., V.S. and E.C. wrote the manuscript
and supplementary information.

\section*{Competing interests}

The authors declare no competing financial interests. 
\end{document}


\title{SUPPLEMENTARY INFORMATION of High-order parametric generation of
coherent XUV radiation}

\maketitle
In the following, we provide additional experimental results to support
the conclusions in the main text and to demonstrate the robustness
of the parametric process.

\subsection*{XUV signal pressure dependence}

\begin{figure}
\begin{centering}
\includegraphics{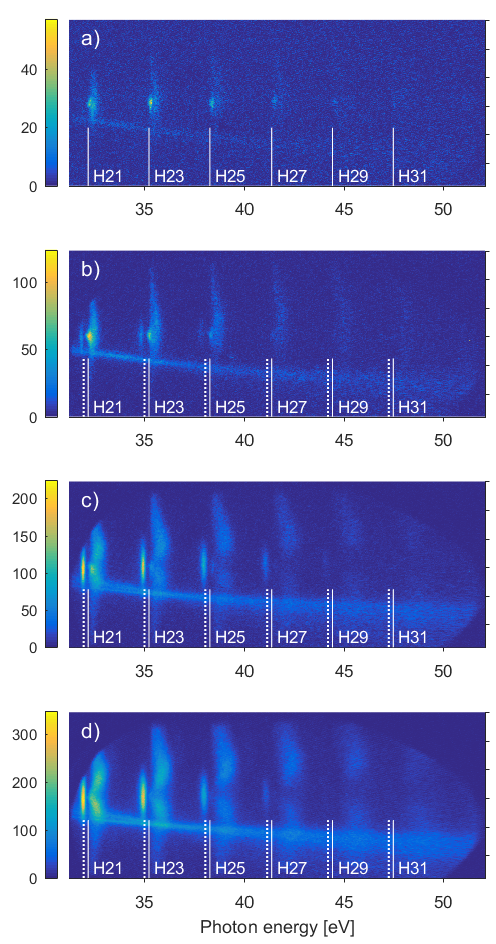}\includegraphics{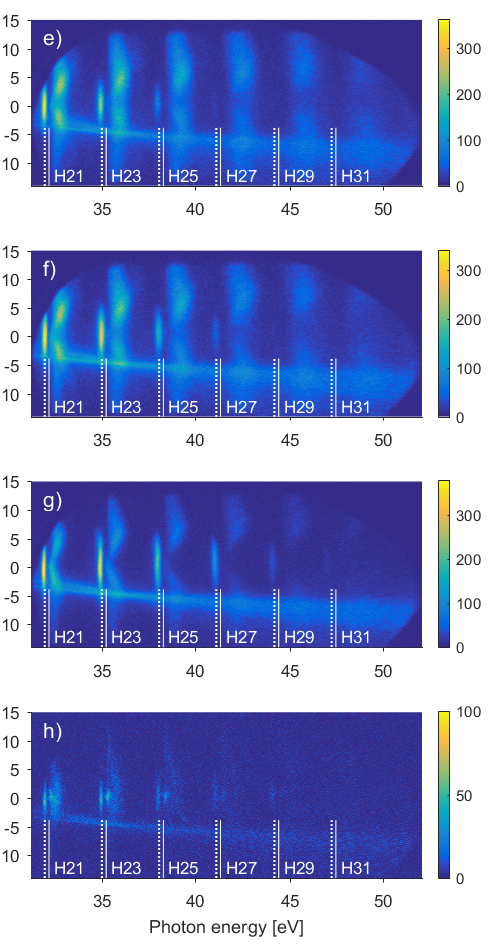}
\par\end{centering}
\caption{\textbf{Experimental spatially resolved XUV spectra generated in krypton
jet.} The y-axis represents the spatial divergence (in mrad) and the
colorbar spatio-spectral intensity (in arbitrary units). The photon
energy $q\omega_{0}$ of the high harmonics are presented by the solid
white lines, while the photon energy of the red-side satellites given
by equation (1) with the parameters $Q=27$ and $m=-2$ are shown
by dashed lines. The driving laser intensity is $\unit[3.5\times10^{15}]{Wcm^{-2}}$
and the pulsed valve opening is delayed by (from a to h) -30, -20,
-10, 0, 10, 20, 30 and $\unit[40]{\mu s}$ to change the gas pressure.
\label{fig:HHG spectra Krypton pressure}}
\end{figure}

The evolution of the XUV spectra as a function of generating medium
pressure was acquired via changing the delay time between the driving
pulse and the nozzle opening time. The zero delay denotes the optimal
time synchronization used for acquisition of data presented in Figs.
1, 2 and 3 in the main text. The positive delay means that the nozzle
opens after the optimal time, meaning that the gas has not yet fully
reached the interaction zone and the driving laser pulse interacts
with lower gas pressure. In the negative delay the laser pulse interacts
with the lower gas pressure because the nozzle is closing and the
gas jet is already spreading into the vacuum chamber. In such a way,
one can effectively and accurately change the gas pressure in the
interaction zone. The experimental results were acquired using the
driving laser intensity $\unit[3.5\times10^{15}]{Wcm^{-2}}$ in the
same conditions as for Fig. 1 in the main text and are presented in
Fig. \ref{fig:HHG spectra Krypton pressure} for several delays.

 Figure \ref{fig:HHG spectra Krypton pressure} presents the evolution
of the HHG and HPG signal in the regime of high intensity, making
the HH peaks very blueshifted and spatially irregular as discussed
in the main text. The HPG signal rises when the time synchronization
approaches the optimal value. However, neither the RSS beam shape
nor the spectral frequency depend on the gas pressure. 

\subsection*{XUV spectra generated in various gases and interaction geometries}

\begin{figure}
\begin{centering}
\includegraphics{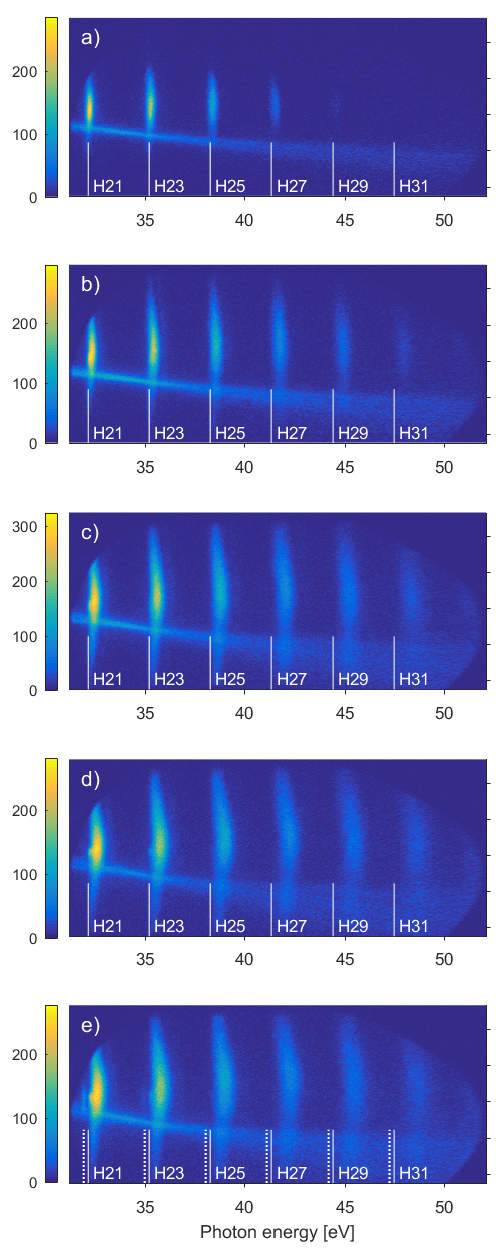}\includegraphics{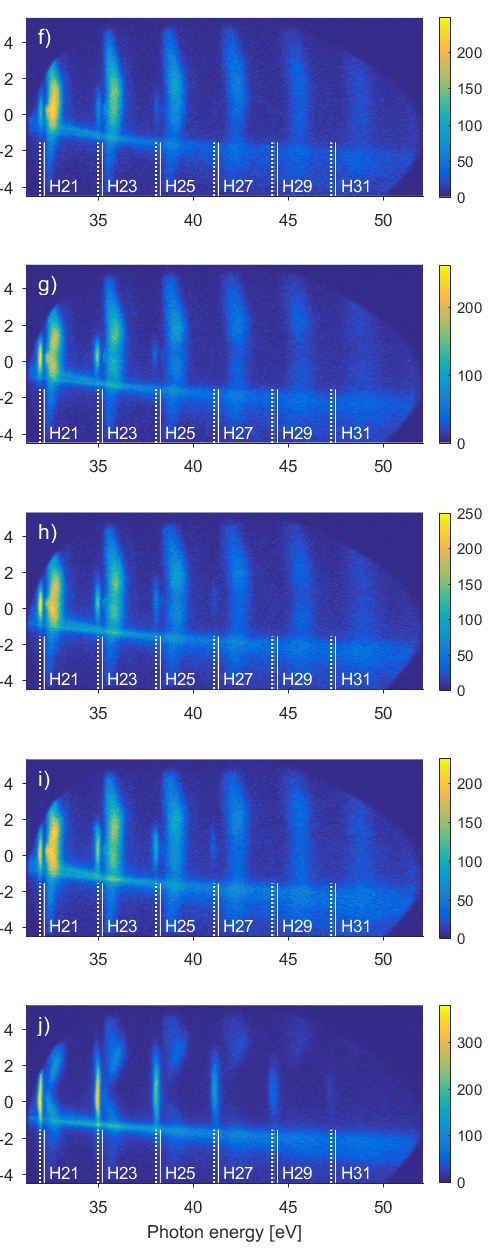}
\par\end{centering}
\caption{\textbf{Experimental spatially resolved XUV spectra generated in a
krypton jet.} The parameters of equation (1) are $Q=27$ and $m=-2$.
The driving laser intensity is (from a to j) 0.15, 0.38, 0.71, 1.1,
1.6, 2.1, 2.5, 2.9, 3.2 and $\unit[3.5\times10^{15}]{Wcm^{-2}}$.\label{fig:Krypton_energy_maintext}}

\end{figure}

\begin{figure}
\begin{centering}
\includegraphics{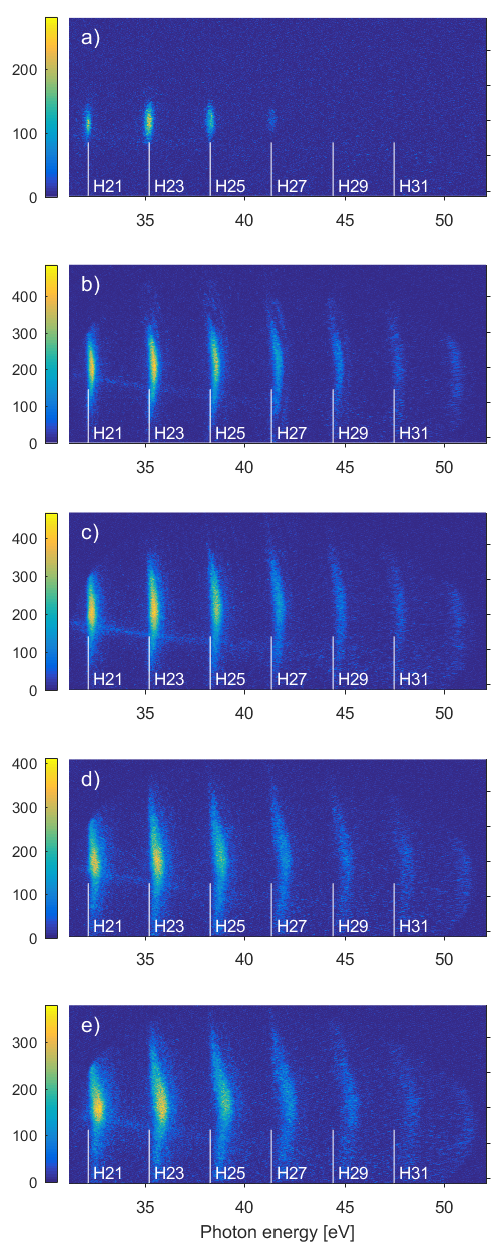}\includegraphics{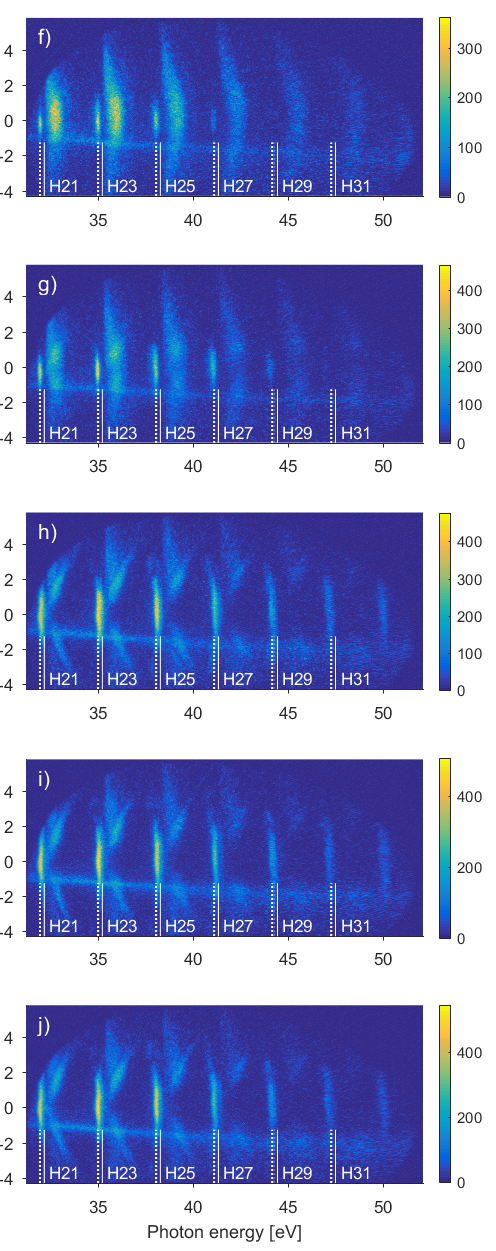}
\par\end{centering}
\caption{\textbf{Experimental spatially resolved XUV spectra generated in argon
jet.} The parameters of equation (1) are $Q=27$ and $m=-2$. The
driving laser intensity is (from a to j) 0.09, 0.15, 0.25, 0.38, 0.53,
0.91, 1.1, 1.6, 1.8 and $\unit[2.3\times10^{15}]{Wcm^{-2}}$.\label{fig: Argon iris-22-focus}}
\end{figure}

\begin{figure}
\begin{centering}
\includegraphics{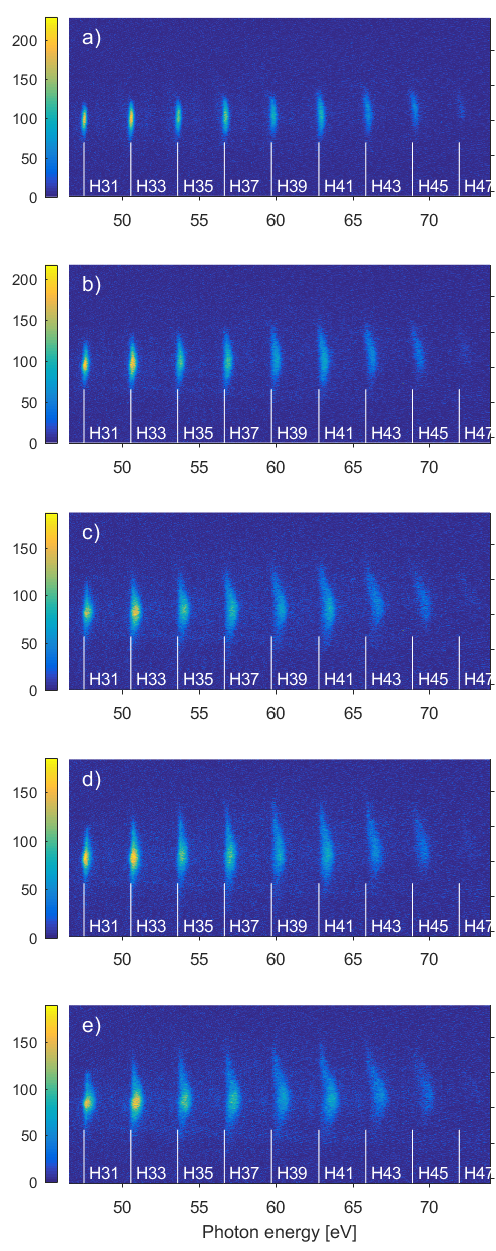}\includegraphics{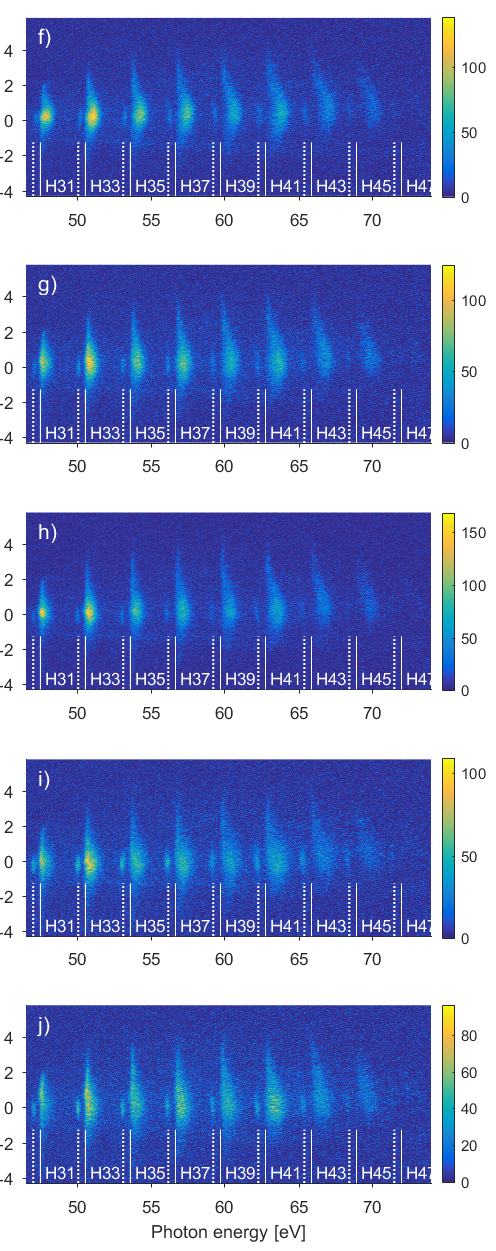}
\par\end{centering}
\caption{\textbf{Experimental spatially resolved XUV spectra generated in neon
jet.} The parameters of equation (1) are $Q=25$ and $m=-4$. The
driving laser intensity is (from a to j) 0.38, 0.71, 1.1, 1.6, 2.1,
2.5, 2.9, 3.2, 3.4 and $\unit[3.5\times10^{15}]{Wcm^{-2}}$.\label{fig:Neon in focus Iris 22 mm}}
\end{figure}

\begin{figure}
\begin{centering}
\includegraphics{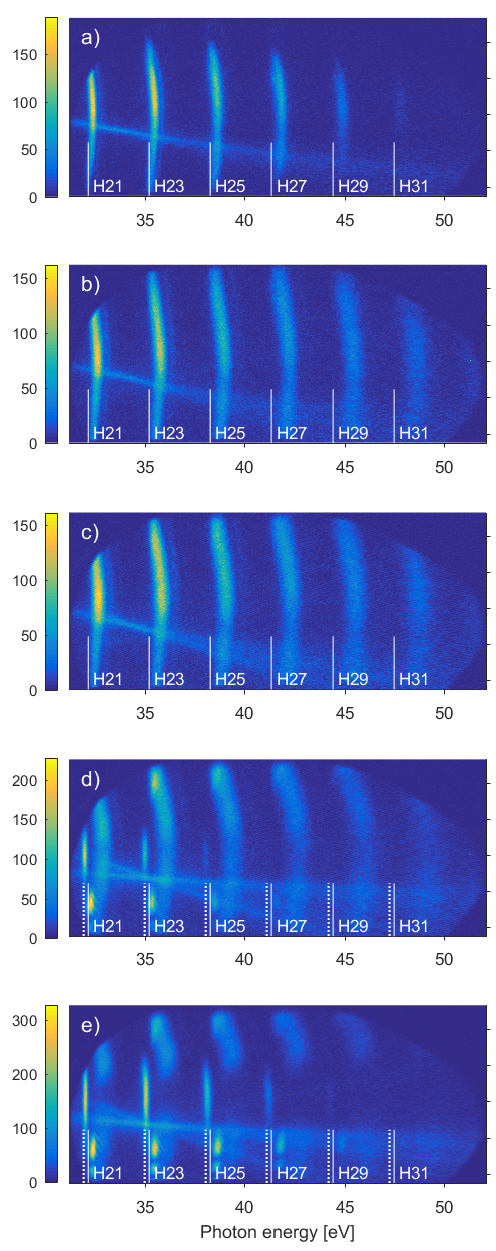}\includegraphics{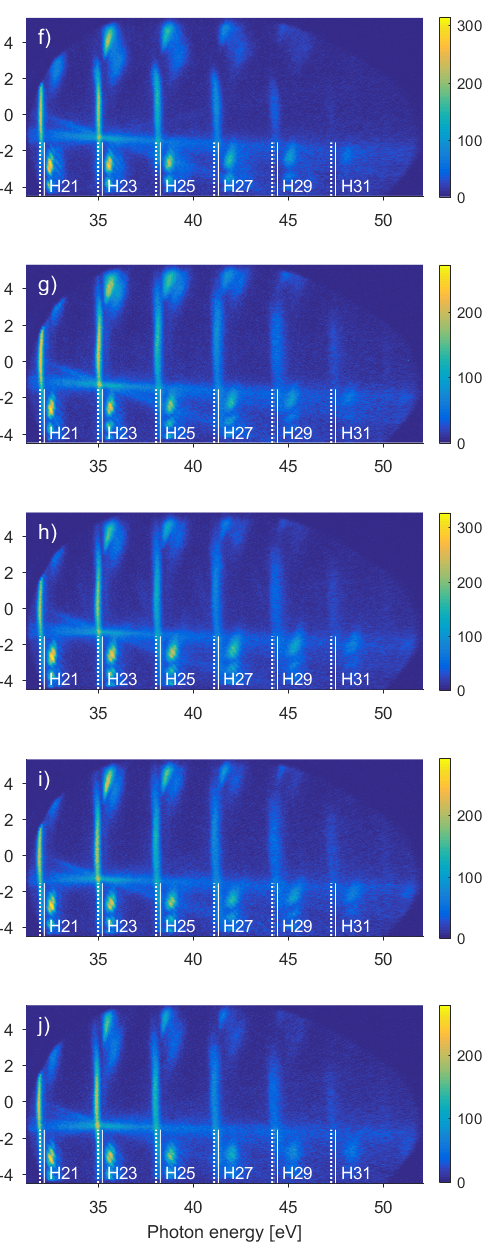}
\par\end{centering}
\caption{\textbf{Experimental spatially resolved XUV spectra generated in krypton
jet located 15 mm after the focal spot.} The parameters of equation
(1) are $Q=27$ and $m=-2$. The driving laser intensity is (from
a to j) 0.12, 0.29, 0.55, 0.87, 1.2, 1.6, 2.0, 2.5, 2.65 and $\unit[2.7\times10^{15}]{Wcm^{-2}}$.
\label{fig: Krypton 15 mm after focus Iris-=00003D 22}}
\end{figure}

\begin{figure}
\begin{centering}
\includegraphics{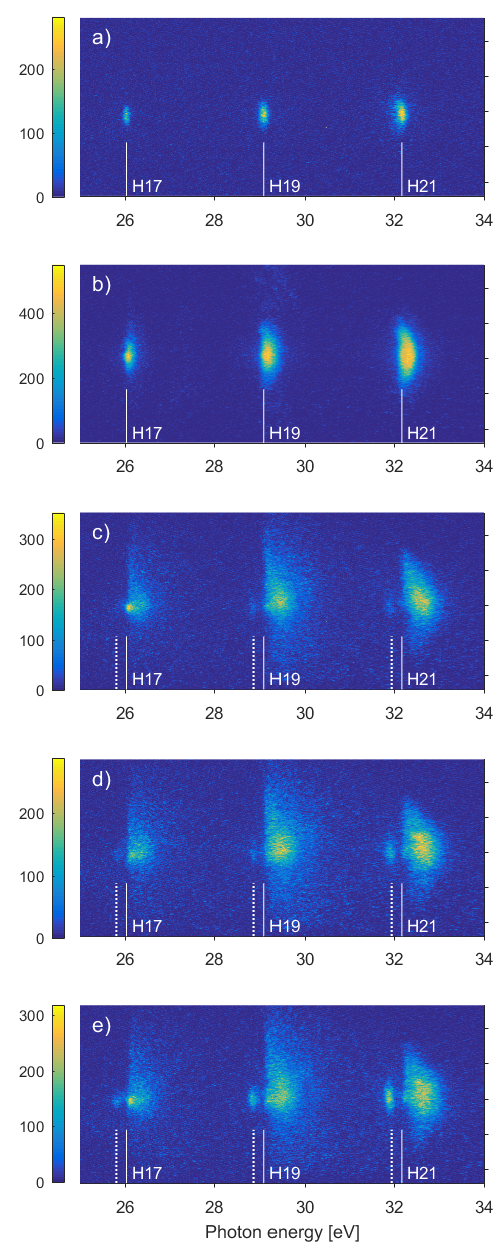}\includegraphics{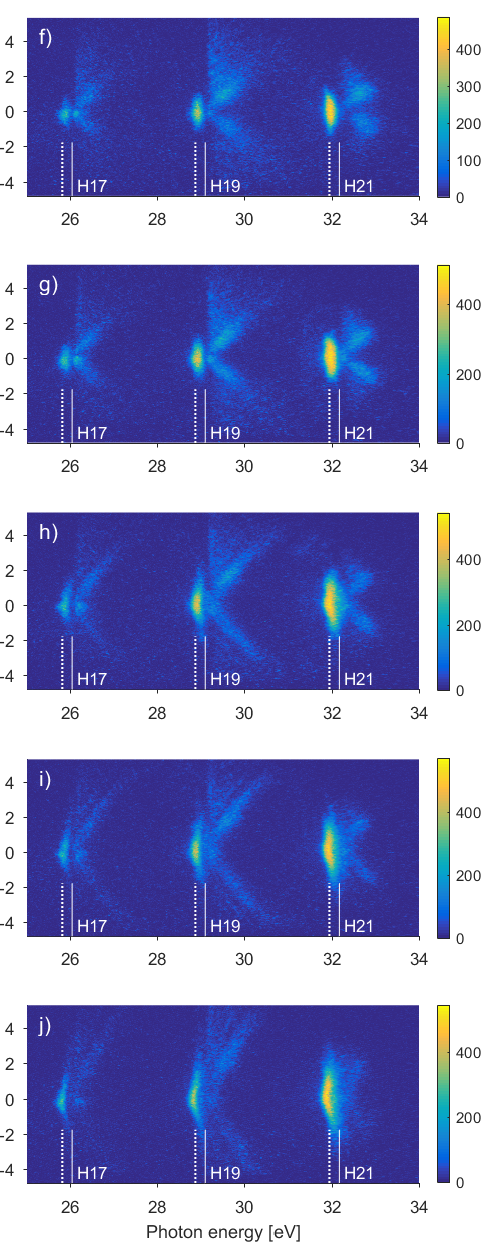}
\par\end{centering}
\caption{\textbf{Experimental spatially resolved XUV spectra generated in argon
jet.} The parameters of equation (1) are $Q=27$ and $m=-2$. The
driving laser intensity is (from a to j) 0.09, 0.25, 0.71, 0.91, 1.1,
1.4, 1.8, 2.5, 2.7 and $\unit[3.1\times10^{15}]{Wcm^{-2}}$.\label{fig: Argon iris-22-focus H17-H21}}
\end{figure}

The Figs. \ref{fig:Krypton_energy_maintext} - \ref{fig: Argon iris-22-focus H17-H21}
present the experimental spatially resolved XUV spectra generated
in various gases and interaction geometries to demonstrate the robustness
of the HPG signal in different experimental conditions. The y-axis
represents the spatial divergence in mrad and the colorbar indicates
spatio-spectral intensity (in arbitrary units). The photon energy
$q\omega_{0}$ of the high harmonics are presented by the solid white
lines, while the photon energy of the red-side satellites given by
equation (1) are shown by dashed lines.

We present XUV spectra generated in krypton, argon and neon jets.
Figures \ref{fig:Krypton_energy_maintext} - \ref{fig: Krypton 15 mm after focus Iris-=00003D 22}
present only harmonic orders of 21 - 47 while similar characteristics
of the XUV radiation are shown for low orders of 17 - 21 in Fig. \ref{fig: Argon iris-22-focus H17-H21}.
Higher orders than 47 are not directly detectable due to the geometry
of the XUV spectrometer.

Figure \ref{fig:Krypton_energy_maintext} expands the dataset shown
on Fig. 1 in the main text comparing the spatially resolved XUV spectra
for more values of driving laser intensity.

Figure \ref{fig: Argon iris-22-focus} shows the XUV  spectrograms
generated in argon jet in similar experimental conditions to Fig.
\ref{fig:Krypton_energy_maintext} and Fig. 1. in the main text but
with lower driving laser intensity.

Figure \ref{fig:Neon in focus Iris 22 mm} shows the XUV spectrograms
generated in neon jet in the same experimental conditions as Fig.
\ref{fig:Krypton_energy_maintext}.

Figure \ref{fig: Krypton 15 mm after focus Iris-=00003D 22} shows
the XUV spectrograms generated in krypton jet with lower driving intensity
corresponding to the gas jet position 15 mm after the focal spot.
Other experimental conditions are similar to those in Fig. \ref{fig:Krypton_energy_maintext}.

Figure \ref{fig: Argon iris-22-focus H17-H21} shows the XUV spectrograms
generated in argon jet in similar experimental conditions to Fig.
\ref{fig:Krypton_energy_maintext}. but the spectrometer was set to
detect harmonic orders 17 - 21. Therefore, it is complementary spectral
range to Fig. \ref{fig: Argon iris-22-focus}.

Overall, these observations show that the effect is robust and occurs
in many different experimental conditions, once the laser intensity
is high enough.